\documentclass[pdflatex,sn-mathphys]{sn-jnl}

\usepackage{amsmath,amssymb,amsfonts}
\usepackage{multirow}
\usepackage{gensymb}
\usepackage{siunitx}

\jyear{2021}%

\theoremstyle{thmstyleone}%
\theoremstyle{thmstyletwo}%

\theoremstyle{thmstylethree}%

\begin{document}

\title[Calibrating the Dice loss for biomedical image segmentation]{Calibrating the Dice loss to handle neural network overconfidence for biomedical image segmentation}


\author*[1,2,3]{\fnm{Michael} \sur{Yeung}}\email{mjyy2@cam.ac.uk}

\author[2,4,5]{\fnm{Leonardo} \sur{Rundo}}\email{lr495@cam.ac.uk}

\author[3]{\fnm{Yang} \sur{Nan}}\email{y.nan20@imperial.ac.uk}

\author[2,4]{\fnm{Evis} \sur{Sala}}\email{es220@medschl.cam.ac.uk}

\author[6]{\fnm{Carola-Bibiane} \sur{Sch\"{o}nlieb}}\email{cbs31@cam.ac.uk}

\author[3]{\fnm{Guang} \sur{Yang}}\email{g.yang@imperial.ac.uk}

\affil*[1]{\orgdiv{School of Clinical Medicine}, \orgname{University of Cambridge}, \orgaddress{\street{Hills Rd}, \city{Cambridge}, \postcode{CB2 0SP}, \country{United Kingdom}}}

\affil[2]{\orgdiv{Department of Radiology}, \orgname{University of Cambridge}, \orgaddress{\street{Hills Rd}, \city{Cambridge}, \postcode{CB2 0QQ}, \country{United Kingdom}}}

\affil[3]{\orgdiv{National Heart \& Lung Institute}, \orgname{Imperial College London}, \orgaddress{\street{Dovehouse St}, \city{London}, \postcode{SW3 6LY}, \country{United Kingdom}}}

\affil[4]{\orgdiv{Cancer Research UK Cambridge Centre}, \orgname{University of Cambridge}, \orgaddress{\street{Robinson Way}, \city{Cambridge}, \postcode{CB2 0RE}, \country{United Kingdom}}}

\affil[5]{\orgdiv{Department of Information and Electrical Engineering and Applied Mathematics (DIEM)}, \orgname{University of Salerno}, \orgaddress{\street{Fisciano}, \city{Salerno}, \postcode{84084}, \country{Italy}}}

\affil[6]{\orgdiv{Department of Applied Mathematics and Theoretical Physics}, \orgname{University of Cambridge}, \orgaddress{\street{Wilberforce Rd}, \city{Cambridge}, \postcode{CB3 0WA}, \country{United Kingdom}}}

\abstract{The Dice similarity coefficient (DSC) is both a widely used metric and loss function for biomedical image segmentation due to its robustness to class imbalance. However, it is well known that the DSC loss is poorly calibrated, resulting in overconfident predictions that cannot be usefully interpreted in biomedical and clinical practice. Performance is often the only metric used to evaluate segmentations produced by deep neural networks, and calibration is often neglected. However, calibration is important for translation into biomedical and clinical practice, providing crucial contextual information to model predictions for interpretation by scientists and clinicians. In this study, we provide a simple yet effective extension of the DSC loss, named the DSC++ loss, that selectively modulates the penalty associated with overconfident, incorrect predictions. As a standalone loss function, the DSC++ loss achieves significantly improved calibration over the conventional DSC loss across six well-validated open-source biomedical imaging datasets, including both 2D binary and 3D multi-class segmentation tasks. Similarly, we observe significantly improved calibration when integrating the DSC++ loss into four DSC-based loss functions. Finally, we use softmax thresholding to illustrate that well calibrated outputs enable tailoring of recall-precision bias, which is an important post-processing technique to adapt the model predictions to suit the biomedical or clinical task. The DSC++ loss overcomes the major limitation of the DSC loss, providing a suitable loss function for training deep learning segmentation models for use in biomedical and clinical practice. Source code is available at: \url{https://github.com/mlyg/DicePlusPlus}}.


\maketitle

\section{Introduction}\label{sec1}

Image segmentation describes a per-pixel classification task, involving partitioning an image into semantic regions based on regional pixel characteristics \cite{pal1993review}. However, class imbalance is frequently observed in biomedical image segmentation tasks, where objects, such as tumours or cell nuclei often occupy a small area relative to the background tissue \cite{roth2015deeporgan}. This can hinder per-pixel classification accuracy and could result in poor segmentation results on biomedical images. To evaluate segmentation quality, the two most popular metrics used are the Dice similarity coefficient (DSC) and the Jaccard Index. Both metrics measure spatial overlap and are therefore generally robust to class imbalance \cite{reinke2021common, fidon2017generalised}.

To incorporate automated image segmentation methods for biomedical applications, not only is segmentation quality important, but it is necessary that predictions are well calibrated \cite{sander2019towards, mehrtash2020confidence, rousseau2021post}. Calibration measures how similar the probabilities assigned to model predictions reflect the real-world underlying uncertainty. In the context of medical image segmentation, a well calibrated model is expected to output predictions with probabilities that match the confidence of an expert human annotator performing manual delineation, or similarly, to match the distribution of segmentations produced by a group of expert annotators.

Importantly, even small differences in imaging hardware or image acquisition parameters may lead to a domain shift that could significantly affect neural network performance, and without proper calibration, result in overconfident predictions that could provide false reassurance and cause potential harm \cite{ghafoorian2017transfer}. Calibration also provides crucial contextual information to the corresponding segmentation output, which is useful for guiding clinical decision making, such as planning for surgical resection or image-guided interventions.

The cross entropy (CE) loss is the most widely used loss function for classification, favoured because of its well calibrated prediction outputs, but it is susceptible to class imbalance and regularly underperforms in these situations, particularly when very small segmentation targets are involved \cite{ma2021loss, yeung2021unified}. In contrast, the DSC loss is, similar to its respective evaluation metric, robust to class imbalance and has been successfully applied to a variety of biomedical image segmentation tasks \cite{milletari2016v, sudre2017generalised, eelbode2020optimization}. However, it is well known that optimising the DSC loss results in poorly calibrated, overconfident predictions (Figure 1) \cite{bertels2019optimization, mehrtash2020confidence, bertels2021theoretical}. 

\begin{figure*}[ht!]
    \centering
    \includegraphics[scale=0.25]{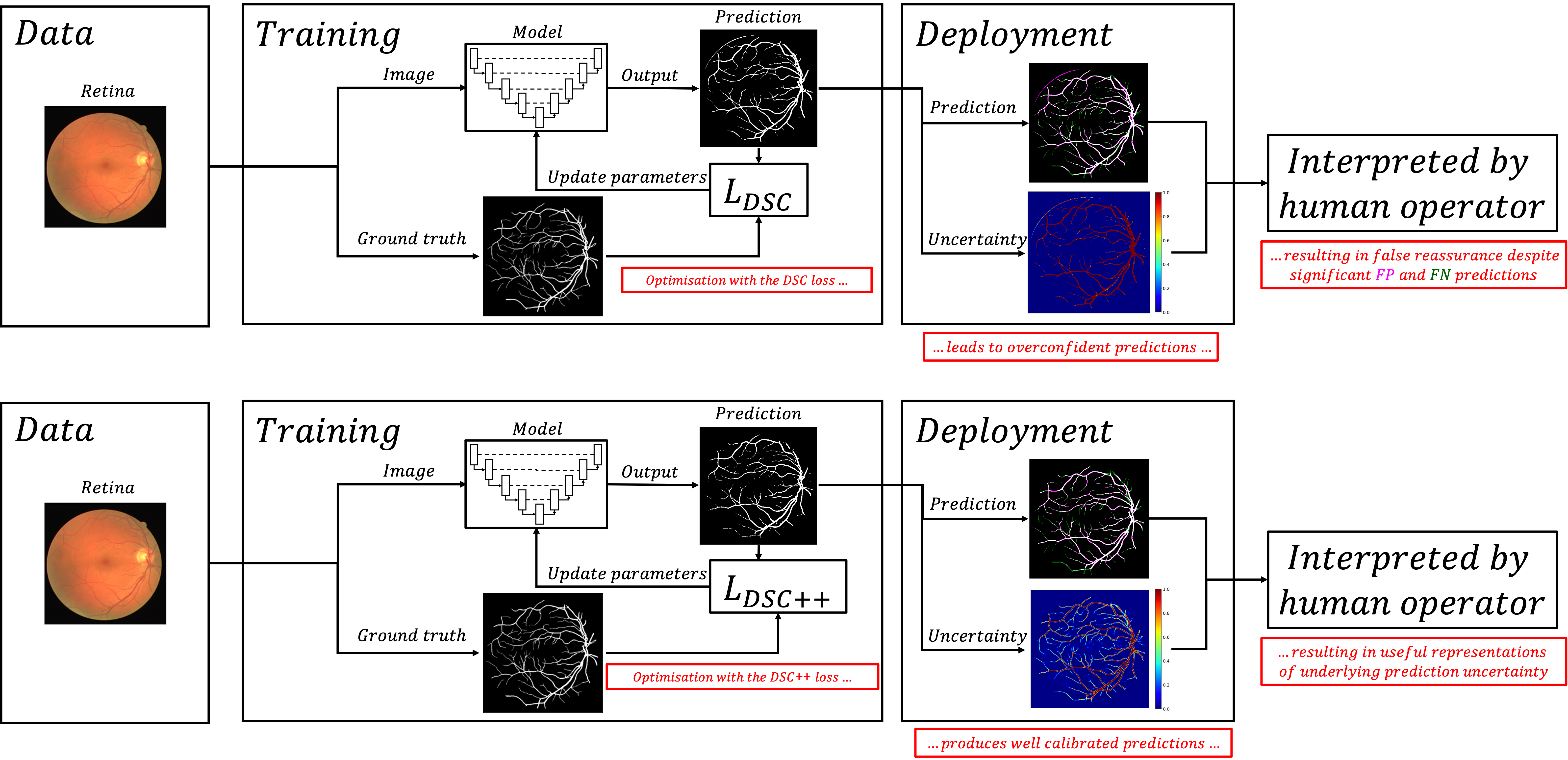}
    \caption{Deep learning-based biomedical image segmentation pipeline. During training, model predictions are compared to ground truth annotations, with model parameters iteratively updated based on the optimisation goal defined by the loss function. During deployment, the model is used for inference, generating a segmentation mask and associated softmax values, which are accessible by the scientist or clinician. Top: Using the DSC loss results in overconfident model predictions, demonstrated by the extreme softmax values illustrated by the heatmap, despite significant false positive (FP) and false negative (FN) predictions. Bottom: In contrast, using the DSC++ loss produces well calibrated predictions that, with a lower certainty, capture the more difficult-to-segment small-diameter retinal vessels. The colours corresponding to the softmax values are shown by the colour-bar on the right.}
    \label{fig:figure_1}
\end{figure*}

The dichotomy between the CE loss, which provides well calibrated but often suboptimal segmentations, and the DSC loss---which produces higher quality segmentations but results in poorly calibrated predictions---suggests that neither loss function is appropriate for clinical use. 

To overcome these challenges in biomedical image segmentation, considerable research has focused on either modifying the CE loss to improve robustness to class imbalance, or improving the calibration of networks trained using the DSC loss. The Focal loss is a variant of the CE loss that addresses the issue of class imbalance by down-weighting the contribution of easy examples enabling learning of harder examples \cite{Lin_2017_ICCV}. Similarly, the exponentially weighted CE loss down-weights correctly predicted samples, but is better suited for smaller degrees of class imbalance \cite{dong2021recognition}. In contrast to directly modifying the CE loss, approaches to improve DSC loss calibration generally focus on modifying the network or applying post hoc calibration. Performing dropout during inference, known as Monte Carlo (MC) dropout, was shown to approximate Bayesian neural networks and improve calibration \cite{gal2016dropout, mehrtash2020confidence, rousseau2021post}. Other network modifications where improved calibration was observed include Platt scaling, which involves fitting a logistic regression model using model outputs, as well as auxiliary networks, which are a generalised version of Platt scaling that instead uses a convolutional layer \cite{platt1999probabilistic, devries2018leveraging}. Avoiding network modifications, deep ensembles involve averaging predictions from multiple, randomly initialised networks, outperforming MC dropout for both performance and calibration \cite{lakshminarayanan2016simple, rousseau2021post, mehrtash2020confidence}. However, ensembling of multiple networks is not only computationally expensive to train, but significantly increases inference time and is therefore of limited use in real-time applications. Finally, improved calibration was observed by initially training a network using the DSC loss, followed by fine-tuning using the CE loss \cite{rousseau2021post}.

Despite various modifications to the CE loss, the segmentation performance remains generally worse than using the DSC loss \cite{yeung2021unified}. In contrast, while the modifications to improve the DSC loss calibration result in comparable calibration to the CE loss, they require pipeline modifications, limiting uptake by the research community as well as clinical applicability. 

The main contributions of this work may be summarised as follows:
\begin{enumerate}
\item We identify the reason for the poor calibration observed with networks trained using the DSC loss, and provide a reformulation, named the DSC++ loss, which directly addresses the issue.
\item We demonstrate significantly improved calibration using the DSC++ loss over the DSC loss, measured using the negative log likelihood (NLL) and Brier score, across six well-validated open-source datasets, including 2D binary and 3D multi-class segmentation tasks.
\item We demonstrate that the DSC++ loss may be readily incorporated to improve the calibration of other DSC-based loss functions.
\item We combine softmax thresholding with the DSC++ loss to enable tailoring of the recall-precision bias for the biomedical or clinical task.
\end{enumerate}

\section{Material and methods}

In this section, we first introduce the CE loss and its variant, the Focal loss, followed by the DSC loss. We then identify the cause of the poor calibration using the DSC loss, and use this to derive the DSC++ loss. After introducing softmax thresholding, the section finally concludes with details on the experimental setup and implementation.

\subsection{CE loss}

CE measures the difference between two probability distributions $y$ and $p$. The CE loss is among the most widely used loss function in machine learning, and in the context of image segmentation, $y$ and $p$ represent the true and predicted distributions over class labels for a given pixel, respectively. The CE loss, ($\mathcal{L}_\text{CE}$), is defined as:

\begin{equation}
\mathcal{L}_\text{CE}(y, p)=-\frac{1}{N} \sum_{i=1}^{N} \sum_{c=1}^{C} y_{i, c} \cdot \log \left(p_{i, c}\right),
\label{eq:CE}
\end{equation}
where $y_{i, c}$ uses a one-hot encoding scheme corresponding to the ground truth labels, $p_{i, c}$ is a matrix of predicted values generated by the model for each class, and where indices $i$ and $c$ iterate over all pixels and classes, respectively. The CE loss is a strictly proper scoring rule, superficially equivalent to the NLL, and therefore yields consistent probabilistic predictions \cite{gneiting2007strictly}.

\subsection{Focal loss}

The Focal loss ($\mathcal{L}_\text{F}$) is an extension of the cross entropy loss developed to addresses the issue of class imbalance in classification tasks \citep{Lin_2017_ICCV}.

The Focal loss uses a modulating factor $\gamma$ to reduce the contribution of easy examples to the overall loss:

\begin{equation}
\mathcal{L}_\text{F}=\boldsymbol{\alpha}\left(1-\left(p_{i, c}\right)\right)^{\gamma} \cdot \mathcal{L}_\text{CE},
\label{eq:CF}
\end{equation}
where $\boldsymbol{\alpha}$ is a vector of class weights, $p_{i,c}$ is a matrix of ground truth probabilities for each class, and $\mathcal{L}_\text{CE}$ is the cross entropy loss as defined in Eq.~(\ref{eq:CE}). The Focal loss is equivalent to the cross entropy loss when $\gamma$ = 1.

\subsection{DSC loss}

CE and the Focal loss are based on pixel-wise error, and therefore in class imbalanced situations, using the CE-based losses result in over-representation of larger objects in the loss, and consequently under-segmentation of smaller objects. Often the segmentation target in biomedical imaging tasks occupies a small area relative to the size of the image, limiting its use as a segmentation quality metric or loss function \cite{yeung2021unified}.

In contrast, the DSC is a spatial overlap index and is therefore robust to class imbalance, and is defined as:

\begin{equation}
\text{DSC} =\frac{1}{C} \sum_{c=1}^{C} \frac{2 \sum_{i=1}^{N} p_{i,c} y_{i,c}}{\sum_{i=1}^{N} p_{i,c}+\sum_{i=1}^{N} y_{i,c}},
\label{eq:DSC}
\end{equation}
where the DSC loss ($\mathcal{L}_\text{DSC}$) is:

\begin{equation}
\mathcal{L}_{\text{DSC}}=1-\text{DSC}.
\label{eq:L_DSC}
\end{equation}

\subsection{DSC++ loss}

The optimisation goal, for both the CE and the DSC loss, is for the neural network to produce confident, and correct, predictions matching the ground truth label. However, neural network overconfidence is a well known phenomenon associated with the DSC loss, but not with the CE loss.
To understand this difference, we provide an equivalent definition of the DSC loss $\mathcal{L}_\text{DSC}$ (Eq.~(\ref{eq:L_DSC})), in terms of true positive (TP), false negative (FN) and false positive predictions (FP):

\begin{equation}
\mathcal{L}_{\text{DSC}}=1-\frac{2 \text{TP}}{2 \text{TP} + \text{FP}+ \text{FN}},
\label{eq:L_DSC_2}
\end{equation}

noting that the DSC score is the harmonic mean of precision and recall, where:

\begin{equation}
\text {Recall}=\frac{\text{TP}}{\text{TP}+\text{FN}},
\label{eq:recall}
\end{equation}

\begin{equation}
\text {Precision}=\frac{\text{TP}}{\text{TP}+\text{FP}}.
\label{eq:precision}
\end{equation}

When both classes are present in equal frequency, the errors associated with the FP and FN predictions are not biased towards a particular class. However, in class imbalanced situations, high precision, low recall solutions are favoured, with over-prediction of the dominant class \cite{salehi2017tversky}. Combined with an optimisation goal that favours confident predictions, this results in networks producing extremely confident, and often incorrect, predictions of the dominant class in regions of uncertainty. 

To overcome this issue, we reformulate the DSC loss to more heavily penalise overconfident predictions. First, we define another equivalent formulation of the $\mathcal{L}_\text{DSC}$, identical in structure to Eq.~(\ref{eq:L_DSC_2}):

\begin{equation}
\mathcal{L}_\text{DSC} = 1-
\frac{1}{C} \sum_{c=1}^{C} \frac{2\sum_{i=1}^{N} p_{0 i,c} y_{0 i,c}}{2\sum_{i=1}^{N} p_{0 i,c} y_{0 i,c}+\sum_{i=1}^{N} p_{0 i,c} y_{1 i,c}+ \sum_{i=1}^{N} p_{1 i,c} y_{0 i,c}},
\label{eq:Tversky_index_2}
\end{equation}
where $p_{0 i,c}$ is the probability of pixel $i$ belonging to the class c, and $p_{1 i,c}$ is the probability of pixel not belonging to class c. Similarly, $y_{0 i}$ is $1$ for class c and $0$ for all other classes, and conversely $y_{1 i}$ takes values of $0$ for class c and $1$ for all other classes.

To penalise overconfidence for uncertain regions, we apply the focal parameter, $\gamma$, directly to the FP and FN predictions, defining the DSC++ loss ($\mathcal{L}_\text{DSC++}$):

\begin{equation}
\mathcal{L}_\text{DSC++} = 1-
\frac{1}{C} \sum_{c=1}^{C}  \frac{2\sum_{i=1}^{N} p_{0 i,c} y_{0 i,c}}{2\sum_{i=1}^{N} p_{0 i,c} y_{0 i,c}+\sum_{i=1}^{N} {(p_{0 i,c} y_{1 i,c})}^\gamma+ \sum_{i=1}^{N} {(p_{1 i,c} y_{0 i,c})}^\gamma}.
\label{eq:Tversky_index_3}
\end{equation}

The DSC++ loss achieves selective penalisation of the overconfident predictions by transforming the penalty from a linear to an exponentially-weighted penalty. When $\gamma = 1$, the DSC++ loss is identical to the DSC loss. When $\gamma > 1$, overconfident predictions are more heavily penalised, with increasing values of $\gamma$ resulting in successively larger penalties applied. Higher $\gamma$ values therefore favour low confidence predictions. The optimal $\gamma$ value balances the maintenance of confident, correct predictions while simultaneously suppressing confident but incorrect predictions.

\subsection{Softmax thresholding}

While the softmax function is not a proxy for uncertainty, the distribution of well calibrated softmax outputs are closely related to the underlying uncertainty, even for out-of-distribution data \cite{guo2017calibration, pearce2021understanding}. To generate a class labelled segmentation output, the argmax function assigns each pixel with the associated class based on the highest softmax value. Rather than using the argmax function, we use a variable threshold that enables manual adjustment of model outputs to favour either precision or recall. Here, we define the output of a model using an indicator function, describing a per-pixel operation that compares the softmax output for the segmentation target, $s$, to a given softmax threshold $\mathcal{T}$:

\begin{equation}
I_{s}=\begin{cases}
1 & \text { if } s<\mathcal{T} \\
0 & \text { otherwise } 
\end{cases}.
\label{eq:softmax_threshold}
\end{equation}

With this generalisation, the argmax function may be restated as a special case where $\mathcal{T} = 0.5$. Higher values of $\mathcal{T}$ favour precision, while lower values favour recall. 

\subsection{Dataset descriptions and evaluation metrics}

To evaluate our proposed loss function, we select six public, well-validated biomedical image segmentation datasets. For retinal vessel segmentation, we use the Digital Retinal Images for Vessel Extraction (DRIVE) dataset \cite{staal2004ridge}. The DRIVE dataset consists of 40 coloured fundus photographs obtained from diabetic retinopathy screening in the Netherlands, with an image resolution of $768 \times 584$ pixels. The Breast UltraSound 2017 (BUS2017) dataset consists of 163 ultrasound images of breast lesions with an average image size of $760 \times 570$ pixels collected from the UDIAT Diagnostic Centre of the Parc Taulí Corporation, Sabadell, Spain \cite{yap2017automated}. Furthermore, we include the 2018 Data Science Bowl (2018DSB) dataset, which contains 670 light microscopy images for nuclei segmentation \cite{caicedo2019nucleus}. For skin lesion segmentation, we use the ISIC2018: Skin Lesion Analysis Towards Melanoma Detection grand challenge dataset. This dataset contains 2,594 images of skin lesions with an average size of $2166 \times 3188$ pixels \cite{codella2019skin}. For colorectal polyp segmentation, we use the CVC-ClinicDB dataset, which consists of 612 frames containing polyps with image resolution 288×368 pixels, generated from 23 video sequences from 13 different patients using standard colonoscopy interventions with white light \cite{bernal2015wm}. Finally, for 3D multi-class segmentation, we use the Kidney Tumour Segmentation 2019 (KiTS19) dataset \citep{heller2019kits19}. This dataset contains 300 arterial phase abdominal CT scans, with voxel-level kidney and kidney tumour annotations. We exclude the 90 scans without associated segmentation masks, and further exclude another 6 scans (case 15, 23, 37, 68, 125 and 133) due to issues with the ground truth quality \citep{heller2019state}.

For all the experiments, except for the DRIVE dataset, which is already partitioned into 20 training and 20 test images, we randomly partitioned the other five datasets into 80\% development and 20\% test set. For all datasets, we further partitioned the development set into 80\% training set and 20\% validation set. Except for the CVC-ClinicDB and KiTS19 datasets, image resolutions are downsampled using bilinear interpolation. For KiTS19, we performed on-the-fly random sampling of patch size $80 \times 160 \times 160$, with patch-wise overlap of $40 \times 80 \times 80$. A summary of the datasets, image resolutions and data partitions are presented in Table 1.

\begin{table*}[ht]
\centering
\caption{Summary of the dataset details and training setup used in these experiments. For KiTS19, the image resolution refers to the patch size used for training.}
\scalebox{0.7}{
\begin{tabular}{l|c|c|c|c|c|c}
\hline
Dataset      & Segmentation         & \#Images & Image resolution & \#Training & \#Validation & \#Test \\ \hline
DRIVE        & Retinal vessel       & 40       & $512 \times 512$        & 16         & 4            & 20     \\
BUS2017      & Breast tumour        & 163      & $128 \times 128$        & 104        & 26           & 33     \\
2018DSB      & Cell nucleus         & 670      & $256 \times 256$        & 428        & 108          & 134    \\
ISIC2018     & Skin lesion          & 2596     & $512 \times 512$        & 1661       & 417          & 518    \\
CVC-ClinicDB & Colorectal polyp     & 612      & $288 \times 384$        & 392        & 98           & 122    \\
KiTS19       & Kidney/Kidney tumour & 204      & $80 \times 160 \times 160$       & 130        & 33           & 41     \\ \hline
\end{tabular}}
\end{table*}

To assess the loss functions, we select two evaluation metrics each for calibration and performance. For calibration, we use the NLL and Brier score, both strictly proper scoring rules. The NLL is equivalent to the CE loss in Eq.~(\ref{eq:CE}), while the Brier score (Brier) computes the mean squared error between predicted probability scores and the true class labels:

\begin{equation}
\mathrm{Brier} =\frac{1}{C}\frac{1}{N} \sum_{i=1}^{C} \sum_{i=1}^{N}(y_{i} - p_{i})^{2}.
\end{equation}

For both metrics, a lower score corresponds to better calibration.

For performance, we use the DSC as previously defined, and the Intersection over Union (IoU), also known as the Jaccard Index:

\begin{equation}
\mathrm{Jaccard}=\frac{\mathrm{TP}}{\mathrm{TP}+\mathrm{FP}+\mathrm{FN}}.
\end{equation}

Contrary to the calibration metrics, a higher DSC or Jaccard score corresponds to better performance.

\subsection{Implementation details}

For our experiments, we leveraged the Medical Image Segmentation with Convolutional Neural Networks (MIScnn) open-source Python library \cite{muller2021miscnn}. This is based on the Keras library using the Tensorflow backend, and all experiments were carried out using NVIDIA P100 GPUs. 

Images were resized as described previously and normalised per-image using the z-score. We applied on-the-fly data augmentation with probability $0.15$, including: scaling ($0.85-1.25\times$), rotation ($-15\degree$ to $+15\degree$), mirroring (vertical and horizontal axes), elastic deformation ($\alpha \in [0, 900]$ and $\sigma \in [9.0, 13.0]$) and brightness ($0.5-2\times$).

To investigate the effect of altering $\gamma$ on the DSC++ loss, we perform a grid search, evaluating values $\gamma \in [0.5, 5]$.

To evaluate the loss functions, we trained the standard U-Net, with model parameters initialised using the Xavier initialisation \cite{ronneberger2015u}. We trained each model with instance normalisation, using the stochastic gradient descent optimiser with a batch size of 1 and initial learning rate of $0.1$ \cite{zhou2019normalization}. For convergence criteria, we used ReduceLROnPlateau to reduce the learning rate by $0.1$ if the validation loss did not improve after 25 epochs, and the EarlyStopping callback to terminate training if the validation loss did not improve after 50 epochs. To compromise for the large patch size used for training on the KiTS19 dataset, we used a stricter convergence criteria of 5 epochs and 10 epochs for the ReduceLROnPlateau and EarlyStopping callbacks respectively.

To evaluate the effect of substituting the DSC loss for the DSC++ loss in several DSC-based variants commonly used to achieve state-of-the-art results, we selected the Tversky loss, Focal Tversky loss, Combo loss and Unified Focal loss \cite{salehi2017tversky,abraham2019novel, taghanaki2019combo, yeung2021unified}. 

The Combo loss ($\mathcal{L}_{\text{Combo}}$) is a compound loss function defined as the weighted sum of the DSC and modified CE loss ($\mathcal{L}_{\text{mCE}}$) \cite{taghanaki2019combo}:

\begin{equation}
\mathcal{L}_\text{Combo}=\alpha\left(\mathcal{L}_\text{mCE}\right)-(1-\alpha) \cdot {\text{DSC}},
\label{eq:combo}
\end{equation}
where: 
\begin{equation}
\mathcal{L}_\text{mCE}=-\frac{1}{N} \sum_{i=1}^{N} \beta\left(y_{i}\ln \left(p_{i}\right)\right)+(1-\beta)\left[\left(1-y_{i}\right) \ln \left(1-p_{i}\right)\right].
\label{eq:MCE}
\end{equation}
The parameters $\alpha$ and $\beta$ take values in the range $[0, 1]$, controlling the relative contribution of the DSC and CE terms to the loss, and the relative weights assigned to false positives and negatives, respectively. Optimising models with the Combo loss has been observed to improve performance, as well as produce visually more consistent segmentations over models trained using the component losses \cite{drozdzal2016importance}.

To overcome the high precision, low recall bias associated with the DSC loss, the Tversky loss ($\mathcal{L}_{\text{Tversky}}$) modifies the weights associated with the FP and FN predictions \cite{salehi2017tversky}:

\begin{equation}
\mathcal{L}_\text{Tversky} = \sum_{c=1}^{C}(1 - TI),
\label{eq:Tversky_loss}
\end{equation}

where the Tversky index (TI) is defined as:

\begin{equation}
\text{TI} = \frac{\sum_{i=1}^{N} p_{0 i} y_{0 i}}{\sum_{i=1}^{N} p_{0 i} y_{0 i}+\alpha\sum_{i=1}^{N} p_{0 i} y_{1 i}+\beta \sum_{i=1}^{N} p_{1 i}, y_{0 i}},
\label{eq:Tversky_index_1}
\end{equation}

where $\alpha$ and $\beta$ control the FP and FN weightings, respectively.

To handle class imbalanced data, the Focal Tversky loss ($\mathcal{L}_\text{FT}$) applies a focal parameter $\gamma$ to alter the weights associated with difficult to classify examples {\cite{abraham2019novel}}:

\begin{equation}
\mathcal{L}_\text{FT}=\sum_{c=1}^{C}(1-\text{TI})^{\frac{1}{\gamma}},
\label{eq:FT}
\end{equation}

$\gamma < 1$ increases the degree of focusing on harder examples.

Finally, the Unified Focal loss ($\mathcal{L}_\text{UF}$) generalises distribution-based and region-based loss functions into a single framework \cite{yeung2021unified}, and is defined as the weighted sum of the Asymmetric Focal loss ($\mathcal{L}_\text{AF}$) and Asymmetric Focal Tversky loss ($\mathcal{L}_\text{AFT}$):

\begin{equation}
\mathcal{L}_\text{UF}=\lambda \mathcal{L}_\text{AF}+(1-\lambda) \mathcal{L}_\text{AFT},
\label{eq:GFL}
\end{equation}
where:

\begin{equation}
\mathcal{L}_\text{AF}=- \frac{\delta}{N} y_{i: r} \log \left(p_{t, r}\right)-\frac{1-\delta}{N} \sum_{c \neq r}\left(1-p_{t, c}\right)^{\gamma} \log \left(p_{t, r}\right),
\end{equation}

\begin{equation}
\mathcal{L}_{\mathrm{AFT}}=\sum_{\mathrm{c} \neq \mathrm{r}}(1-\mathrm{TI})+\sum_{\mathrm{c}=\mathrm{r}}(1-\mathrm{TI})^{1-\gamma},
\end{equation}

where the TI is redefined as:

\begin{equation}
\operatorname{TI} =
\frac{\sum_{i=1}^{N} p_{0 i} y_{0 i}}{\sum_{i=1}^{N} p_{0 i} y_{0 i}+(1-\delta)\sum_{i=1}^{N} p_{0 i} y_{1 i}+\delta \sum_{i=1}^{N} p_{1 i} y_{0 i}}.
\label{eq:modified_Tversky_index}
\end{equation}

The three hyperparameters are $\lambda$, which controls the relative weights of the two component losses, $\delta$, which controls the relative weighting of positive and negative examples, and $\gamma$, which controls the relative weighting of easy and difficult examples. 

We used the optimal hyperparameters as described in the original papers, detailed in Table 2. For each loss function, we substituted the DSC component of the loss for the DSC++ loss, setting $\gamma = 2$.

\begin{table}[ht!]
\centering
\caption{Hyperparameter settings used in these experiments for the DSC and cross entropy-based loss functions.}
\begin{tabular}{l|ccccc}
\hline
              & \multicolumn{5}{c}{Hyperparameter}                                                                              \\ \hline
Loss          & \multicolumn{1}{c|}{$\alpha$}   & \multicolumn{1}{c|}{$\beta$}   & \multicolumn{1}{c|}{$\gamma$}   & \multicolumn{1}{c|}{$\delta$}   & $\lambda$   \\ \hline
Focal         & \multicolumn{1}{c|}{0.5} & \multicolumn{1}{c|}{-}   & \multicolumn{1}{c|}{2}   & \multicolumn{1}{c|}{-}   & -   \\
Tversky       & \multicolumn{1}{c|}{0.3} & \multicolumn{1}{c|}{0.7} & \multicolumn{1}{c|}{-}   & \multicolumn{1}{c|}{-}   & -   \\
Focal Tversky & \multicolumn{1}{c|}{0.3} & \multicolumn{1}{c|}{0.7} & \multicolumn{1}{c|}{$\frac{4}{3}$} & \multicolumn{1}{c|}{-}   & -   \\
Combo         & \multicolumn{1}{c|}{0.5} & \multicolumn{1}{c|}{0.5} & \multicolumn{1}{c|}{-}   & \multicolumn{1}{c|}{-}   & -   \\
Unified Focal & \multicolumn{1}{c|}{-}   & \multicolumn{1}{c|}{-}   & \multicolumn{1}{c|}{0.5} & \multicolumn{1}{c|}{0.6} & 0.5 \\ \hline
\end{tabular}
\end{table}

To test for statistical significance, we used the Wilcoxon rank sum test. A statistically significant difference was defined as $p < 0.05$. We use bootstrapping to calculate the standard errors for each metric. To evaluate effect of softmax thresholding, we selected thresholds $\mathcal{T}\in[0.05,0.95]$ using the DSC and DSC++ loss on the DRIVE dataset.

\section{Results}

In this section, we first describe the results for the hyperparameter experiments using the DSC++ loss, before comparing the DSC loss, CE loss, Focal loss and DSC++ loss on five 2D binary segmentation tasks, followed by on one 3D multi-class segmentation task. Next, we compare the performance and calibration of various DSC-based loss functions with and without the DSC++ modification. Finally, we compare the effects of softmax thresholding using the DSC and DSC++ loss on recall-precision bias.

\subsection{DSC++ loss hyperparameter tuning}

The results for the hyperparameter experiments using the DSC++ loss are shown in Table 3.

\begin{table*}[ht]
\centering
\caption{Calibration and performance of the DSC++ loss on the DRIVE dataset with different $\gamma$ values. The standard errors are shown in brackets. The best scores are denoted in bold.}
\scalebox{0.75}{
\begin{tabular}{c|c|c|cc}
\hline
\multicolumn{1}{l|}{}      & \multicolumn{2}{c|}{Uncertainty}               & \multicolumn{2}{c}{Performance}                                     \\ \hline
\multicolumn{1}{l|}{Gamma} & NLL ($\downarrow$)                         & Brier ($\downarrow$)                       & \multicolumn{1}{c|}{Dice ($\uparrow$)}                         & Jaccard ($\uparrow$)                      \\ \hline
0.5                        & 0.281 ($\pm 0.019$)          & 0.033 ($\pm 0.001$)          & \multicolumn{1}{c|}{0.804 ($\pm 0.003$)}          & 0.673 ($\pm 0.004$)          \\
1.0                          & 0.204 ($\pm 0.014$)          & 0.031 ($\pm 0.001$)          & \multicolumn{1}{c|}{0.804 ($\pm 0.003$)}          & 0.672 ($\pm 0.05$)          \\
1.5                        & 0.067 ($\pm 0.005$)          & 0.026 ($\pm 0.001$)          & \multicolumn{1}{c|}{0.804 ($\pm 0.004$)}          & 0.673 ($\pm 0.005$)          \\
2.0                         & 0.041 ($\pm 0.003$)          & \textbf{0.024 ($\pm 0.001$)} & \multicolumn{1}{c|}{\textbf{0.808 ($\pm 0.003$)}} & \textbf{0.678 ($\pm 0.004$)} \\
2.5                        & 0.038 ($\pm 0.002$)          & \textbf{0.024 ($\pm 0.001$)} & \multicolumn{1}{c|}{0.804 ($\pm 0.003$)}          & 0.673 ($\pm 0.05$)          \\
3.0                          & 0.038 ($\pm 0.002$)          & 0.027 ($\pm 0.001$)          & \multicolumn{1}{c|}{0.797 ($\pm 0.004$)}          & 0.664 ($\pm 0.006$)          \\
3.5                        & \textbf{0.035 ($\pm 0.002$)} & 0.031 ($\pm 0.001$)          & \multicolumn{1}{c|}{0.804 ($\pm 0.004$)}          & 0.672 ($\pm 0.005$)          \\
4.0                          & 0.038 ($\pm 0.002$)          & 0.034 ($\pm 0.001$)          & \multicolumn{1}{c|}{0.796 ($\pm 0.004$)}          & 0.661 ($\pm 0.006$)          \\
4.5                        & 0.039 ($\pm 0.002$)          & 0.042 ($\pm 0.001$)          & \multicolumn{1}{c|}{0.795 ($\pm 0.004$)}          & 0.660 ($\pm 0.005$)          \\
5.0                          & 0.041 ($\pm 0.002$)          & 0.048 ($\pm 0.001$)          & \multicolumn{1}{c|}{0.794 ($\pm 0.004$)}          & 0.658 ($\pm 0.006$)          \\ \hline
\end{tabular}}
\end{table*}

Most noticeable is the significant decrease in the NLL with values of $\gamma > 1$. The NLL decreases with increasing $\gamma$, appearing to plateau at $\gamma = 2$. Similarly, Brier score decreases with increasing $\gamma$, with the lowest Brier scores at $\gamma$ values of $2$ and $2.5$. With $\gamma = 2$, there is a statistically significance difference in NLL ($p=\num{6e-8}$) and Brier ($p=\num{2e-7}$) scores compared to the DSC loss. However, above this range, increasing gamma leads to an increase in NLL and Brier score. In terms of performance metrics, the highest DSC and Jaccard scores were observed with $\gamma = 2$, at $0.808$ and $0.678$ respectively. There is no statistically significant difference between the DSC and Jaccard scores using different $\gamma$ values, suggesting that the improved calibration score does not come at the cost to performance. 

To understand whether $\gamma$ improves calibration scores by reducing model overconfidence, Figure 2 shows an example of the softmax probability outputs for an example test-set image.

\begin{figure*}[ht!]
    \centering
    \includegraphics[scale=0.35]{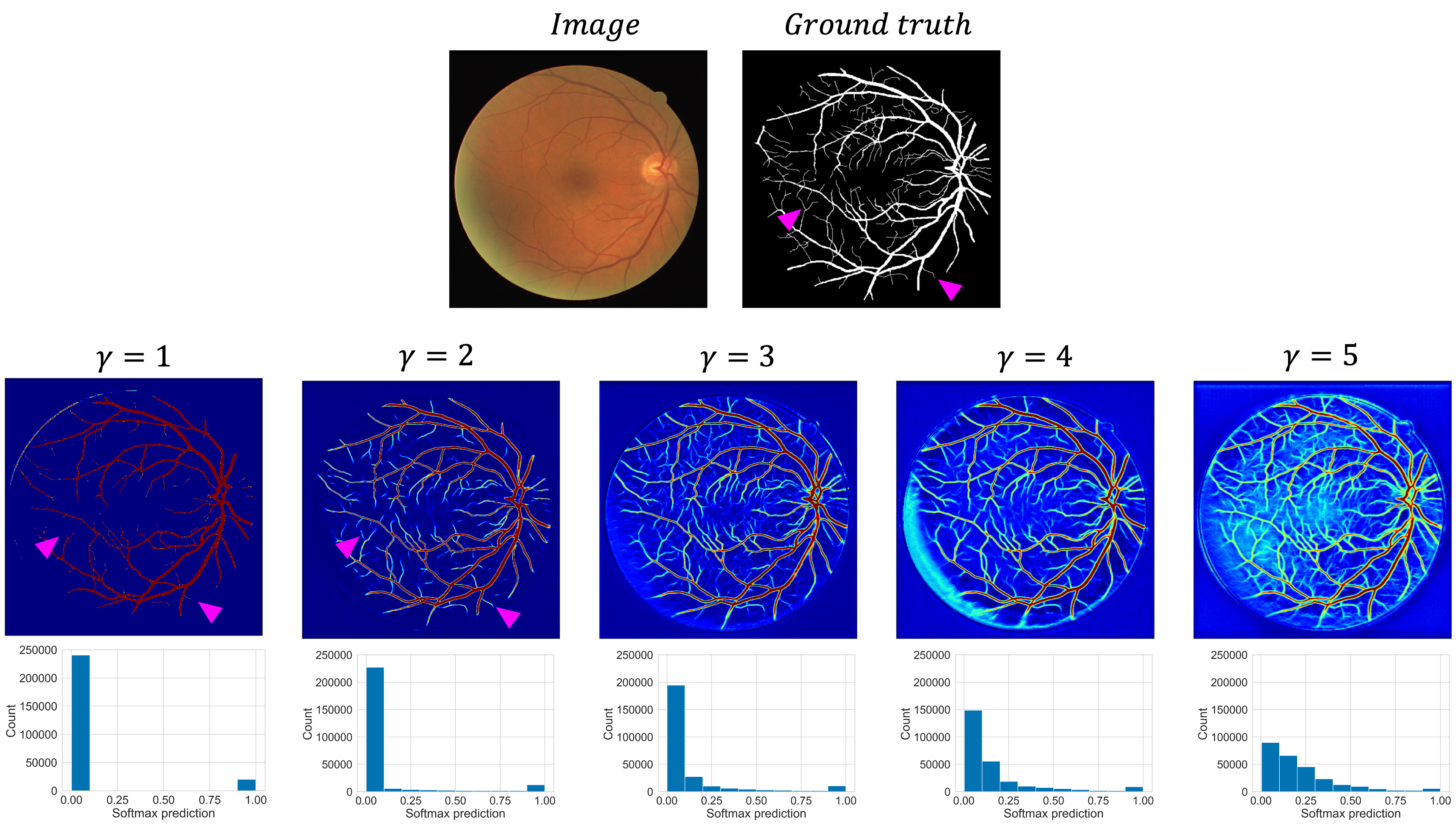}
    \caption{The effect of altering $\gamma$ on the softmax prediction outputs. Top: input image and ground truth segmentation. The pink arrows highlight example areas where segmentation quality differs. Middle: the softmax predictions for each model trained using the DSC++ loss with different $\gamma$ value are displayed as heatmaps. Bottom: Histogram plots showing the softmax predictions and corresponding number of pixels.}
    \label{fig:figure_2}
\end{figure*}

With increasing $\gamma$ values, there is a reduction in overconfident model predictions, in comparison to the DSC loss ($\gamma = 1$), where model predictions are concentrated at the extremes. Importantly, the low confidence areas are concentrated around the difficult to segment smaller retinal vessels, providing a plausible approximation of the underlying uncertainty. 

\subsection{Loss function comparisons}

\subsubsection{2D binary segmentation}

The CE loss, Focal loss, DSC loss and DSC++ loss were evaluated on five, 2D binary biomedical imaging datasets. Based on the results from the hyperparameter investigation, we set $\gamma = 2$ for the DSC++ loss. The results are shown in Table 4. 

\begin{table*}[ht!]
\centering
\caption{Calibration and performance of different loss functions on five biomedical imaging datasets. The standard errors are shown in brackets. The best scores are denoted in bold.}
\scalebox{0.65}{
\begin{tabular}{l|c|cc|cc}
\hline
                              & \multicolumn{1}{l|}{} & \multicolumn{2}{c|}{Calibration}                                     & \multicolumn{2}{c}{Performance}                                      \\ \hline
Dataset                       & Loss                  & \multicolumn{1}{c|}{NLL ($\downarrow$)}                    & Brier ($\downarrow$)                 & \multicolumn{1}{c|}{Dice ($\uparrow$)}                   & Jaccard ($\uparrow$)               \\ \hline
\multirow{4}{*}{Drive}        & CE                    & \multicolumn{1}{c|}{0.051 ($\pm 0.003$)}          & \textbf{0.024 ($\pm 0.001$)} & \multicolumn{1}{c|}{0.798 ($\pm 0.004$)}          & 0.664 ($\pm 0.005$)          \\
                              & Focal                 & \multicolumn{1}{c|}{0.048 ($\pm 0.002$)}          & 0.036 ($\pm 0.001$)          & \multicolumn{1}{c|}{0.795 ($\pm 0.005$)}          & 0.660 ($\pm 0.007$)          \\
                              & DSC                   & \multicolumn{1}{c|}{0.204 ($\pm 0.013$)}          & 0.031 ($\pm 0.001$)          & \multicolumn{1}{c|}{0.804 ($\pm 0.003$)}          & 0.672 ($\pm 0.005$)          \\
                              & DSC++                 & \multicolumn{1}{c|}{\textbf{0.041 ($\pm 0.003$)}} & \textbf{0.024 ($\pm 0.001$)} & \multicolumn{1}{c|}{\textbf{0.808 ($\pm 0.003$)}} & \textbf{0.678 ($\pm 0.004$)} \\ \hline
\multirow{4}{*}{BUS2017}      & CE                    & \multicolumn{1}{c|}{0.020 ($\pm 0.005$)}          & 0.014 ($\pm 0.003$)          & \multicolumn{1}{c|}{0.787 ($\pm 0.037$)}          & 0.690 ($\pm 0.041$)          \\
                              & Focal                 & \multicolumn{1}{c|}{\textbf{0.019 ($\pm 0.004$)}} & 0.020 ($\pm 0.003$)          & \multicolumn{1}{c|}{0.770 ($\pm 0.041$)}          & 0.673 ($\pm 0.042$)          \\
                              & DSC                   & \multicolumn{1}{c|}{0.137 ($\pm 0.046$)}          & 0.022 ($\pm 0.005$)          & \multicolumn{1}{c|}{0.784 ($\pm 0.038$)}          & 0.688 ($\pm 0.042$)          \\
                              & DSC++                 & \multicolumn{1}{c|}{0.034 ($\pm 0.016$)}          & \textbf{0.013 ($\pm 0.004$)} & \multicolumn{1}{c|}{\textbf{0.842 ($\pm 0.031$)}} & \textbf{0.756 ($\pm 0.034$)} \\ \hline
\multirow{4}{*}{2018DSB}      & CE                    & \multicolumn{1}{c|}{\textbf{0.033 ($\pm 0.003$)}} & \textbf{0.019 ($\pm 0.002$)} & \multicolumn{1}{c|}{0.912 ($\pm 0.006$)}          & 0.845 ($\pm 0.009$)          \\
                              & Focal                 & \multicolumn{1}{c|}{0.044 ($\pm 0.004$)}          & 0.028 ($\pm 0.002$)          & \multicolumn{1}{c|}{0.904 ($\pm 0.007$)}          & 0.832 ($\pm 0.010$)          \\
                              & DSC                   & \multicolumn{1}{c|}{0.167 ($\pm 0.019$)}          & 0.025 ($\pm 0.002$)          & \multicolumn{1}{c|}{\textbf{0.916 ($\pm 0.006$)}} & \textbf{0.852 ($\pm 0.009$)} \\
                              & DSC++                 & \multicolumn{1}{c|}{\textbf{0.033 ($\pm 0.004$)}} & \textbf{0.019 ($\pm 0.002$)} & \multicolumn{1}{c|}{\textbf{0.916 ($\pm 0.006$)}} & 0.850 ($\pm 0.009$)          \\ \hline
\multirow{4}{*}{ISIC2018}     & CE                    & \multicolumn{1}{c|}{0.083 ($\pm 0.010$)}          & 0.036 ($\pm 0.003$)          & \multicolumn{1}{c|}{0.863 ($\pm 0.008$)}          & 0.787 ($\pm 0.009$)          \\
                              & Focal                 & \multicolumn{1}{c|}{\textbf{0.068 ($\pm 0.005$)}} & 0.041 ($\pm 0.002$)          & \multicolumn{1}{c|}{0.865 ($\pm 0.008$)}          & 0.793 ($\pm 0.009$)          \\
                              & DSC                   & \multicolumn{1}{c|}{0.373 ($\pm 0.037$)}          & 0.044 ($\pm 0.003$)          & \multicolumn{1}{c|}{\textbf{0.883 ($\pm 0.006$)}} & \textbf{0.812 ($\pm 0.008$)} \\
                              & DSC++                 & \multicolumn{1}{c|}{0.086 ($\pm 0.011$)}          & \textbf{0.034 ($\pm 0.003$)} & \multicolumn{1}{c|}{0.882 ($\pm 0.006$)}          & 0.811 ($\pm 0.008$)          \\ \hline
\multirow{4}{*}{CVC-ClinicDB} & CE                    & \multicolumn{1}{c|}{0.041 ($\pm 0.008$)}          & 0.015 ($\pm 0.002$)          & \multicolumn{1}{c|}{0.870 ($\pm 0.014$)}          & 0.796 ($\pm 0.017$)          \\
                              & Focal                 & \multicolumn{1}{c|}{\textbf{0.028 ($\pm 0.004$)}} & 0.014 ($\pm 0.002$)          & \multicolumn{1}{c|}{0.893 ($\pm 0.013$)}          & 0.828 ($\pm 0.015$)          \\
                              & DSC                   & \multicolumn{1}{c|}{0.167 ($\pm 0.033$)}          & 0.019 ($\pm 0.003$)          & \multicolumn{1}{c|}{0.884 ($\pm 0.014$)}          & 0.817 ($\pm 0.016$)          \\
                              & DSC++                 & \multicolumn{1}{c|}{0.037 ($\pm 0.007$)}          & \textbf{0.013 ($\pm 0.002$)} & \multicolumn{1}{c|}{\textbf{0.894 ($\pm 0.013$)}} & \textbf{0.829 ($\pm 0.015$)} \\ \hline
\end{tabular}}
\end{table*}

Firstly, there was a statistically significant difference between the NLL using DSC++ loss compared to the DSC loss, across all datasets (DRIVE: $p=\num{6e-8}$, BUS2017: $p=0.01$, 2018DSB: $p=\num{8e-13}$, ISIC2018: $p=\num{1e-12}$ and CVC-ClinicDB: $p=\num{2e-11}$). There was no significant difference between the NLL values using the CE, Focal or DSC++ loss.  The DSC++ loss achieved the lowest Brier score for all five datasets, with statistically significant differences observed on the DRIVE ($p=\num{2e-7}$), ISIC2018 ($p=\num{0.01}$) and CVC-ClinicDB ($p=\num{0.04}$) datasets compared to the DSC loss. The DSC++ loss achieved the highest DSC score for four out of the five datasets, and the highest Jaccard score for three out of the five datasets. In contrast, the CE-based loss achieved the lowest performance scores across all datasets, with the Focal loss achieving the lowest Dice and Jaccard score for four out of the five datasets. A statistically significant difference ($p < 0.05$) was observed on the DRIVE dataset for both the DSC and Jaccard scores between the DSC++ and CE-based losses.

Example segmentations using each loss function for the five datasets are shown in Figure 3. Visually, the best segmentations are observed using the DSC++ loss. While model predictions derived from CE-based losses appear well calibrated, the segmentation quality is generally poor. In contrast, the DSC loss, despite very confident predictions, produces considerable false positive predictions such as in the BUS2017 example, as well as false negative predictions as seen in the CVC-ClinicDB example. 

\begin{figure*}[ht!]
    \centering
    \includegraphics[scale=0.35]{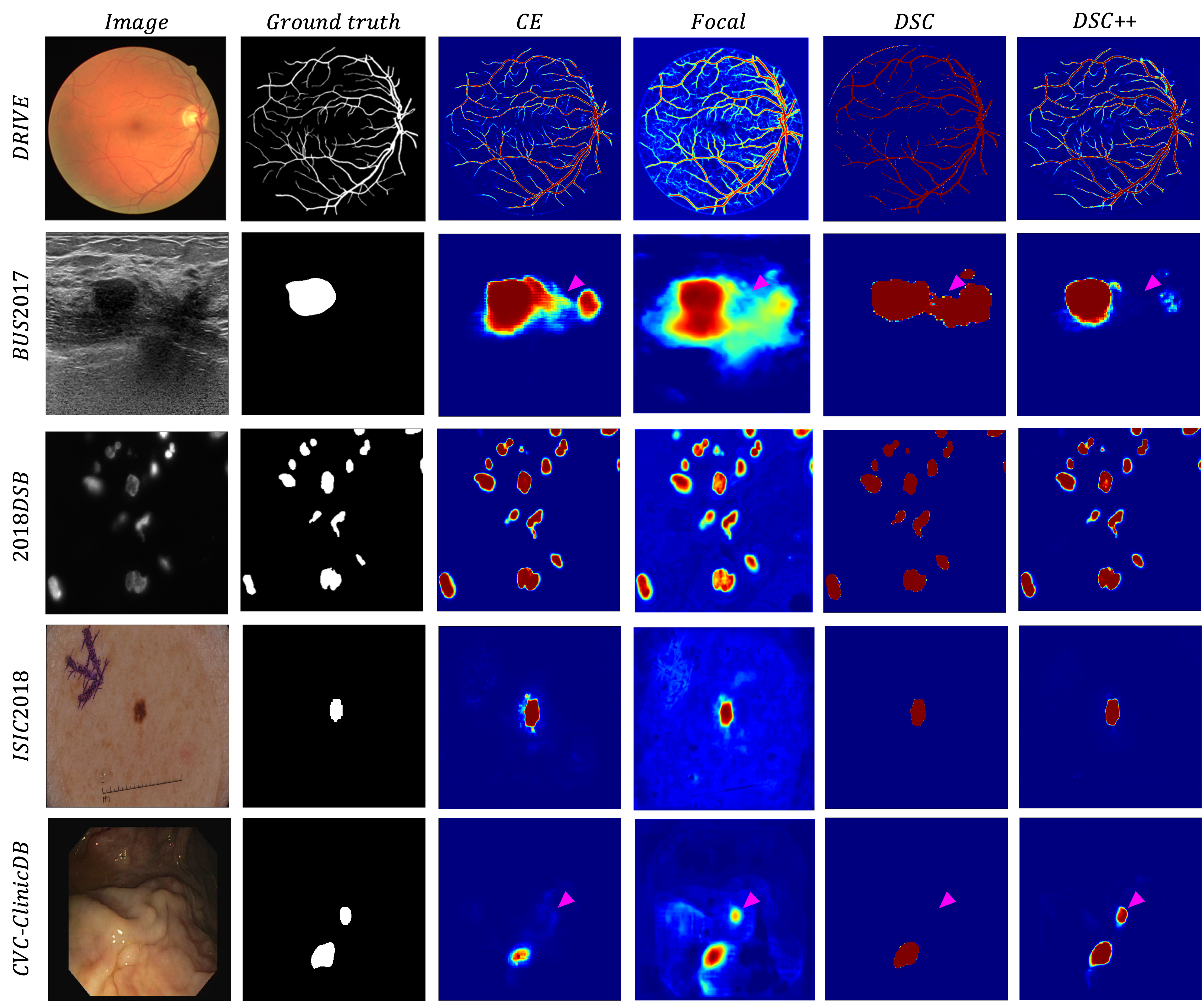}
    \caption{Example segmentations, with softmax predictions visualised as a heatmap, for each loss function for each of the five datasets. The image and ground truth are provided for reference. The pink arrows highlight example areas where segmentation quality differs.}
    \label{fig:figure_3}
\end{figure*}

\subsubsection{3D multi-class segmentation}

The performance of the CE loss, Focal loss, DSC loss and DSC++ loss was further evaluated on the KiTS19 dataset, a 3D multi-class segmentation task. The results are shown in Table 5.

\begin{table*}[ht!]
\centering
\caption{Calibration and performance of different loss functions on the KiTS19 dataset. The standard errors are shown in brackets. The best scores are denoted in bold.}
\scalebox{0.5}{
\begin{tabular}{ccccc|cccc}
\cline{2-9}
\multicolumn{1}{l}{}       & \multicolumn{4}{c|}{Kidney}                                                                                                                                      & \multicolumn{4}{c}{Kidney tumour}                                                                                                                                \\ \hline
\multicolumn{1}{l|}{}      & \multicolumn{2}{c|}{Calibration}                                                          & \multicolumn{2}{c|}{Performance}                                     & \multicolumn{2}{c|}{Calibration}                                                          & \multicolumn{2}{c}{Performance}                                      \\ \hline
\multicolumn{1}{c|}{Loss}  & \multicolumn{1}{c|}{NLL ($\downarrow$)}                    & \multicolumn{1}{c|}{Brier ($\downarrow$)}                  & \multicolumn{1}{c|}{Dice ($\uparrow$)}                   & Jaccard ($\uparrow$)               & \multicolumn{1}{c|}{NLL ($\downarrow$)}                    & \multicolumn{1}{c|}{Brier ($\downarrow$)}                  & \multicolumn{1}{c|}{Dice ($\uparrow$)}                   & Jaccard ($\uparrow$)                \\ \hline
\multicolumn{1}{c|}{CE}    & \multicolumn{1}{c|}{\textbf{0.012 ($\pm 0.005$)}} & \multicolumn{1}{c|}{\textbf{0.007 ($\pm 0.003$)}} & \multicolumn{1}{c|}{0.896 ($\pm 0.012$)}          & 0.819 ($\pm 0.017$)          & \multicolumn{1}{c|}{0.031 ($\pm 0.003$)}          & \multicolumn{1}{c|}{0.012 ($\pm 0.002$)}          & \multicolumn{1}{c|}{0.188 ($\pm 0.035$)}          & 0.124 ($\pm 0.025$)          \\
\multicolumn{1}{c|}{Focal} & \multicolumn{1}{c|}{\textbf{0.012 ($\pm 0.004$)}} & \multicolumn{1}{c|}{0.008 ($\pm 0.003$)}          & \multicolumn{1}{c|}{\textbf{0.911 ($\pm 0.009$)}} & \textbf{0.841 ($\pm 0.014$)} & \multicolumn{1}{c|}{\textbf{0.020 ($\pm 0.002$)}} & \multicolumn{1}{c|}{\textbf{0.011 ($\pm 0.002$)}} & \multicolumn{1}{c|}{0.301 ($\pm 0.043$)}          & 0.213 ($\pm 0.035$)          \\
\multicolumn{1}{c|}{DSC}   & \multicolumn{1}{c|}{0.050 ($\pm 0.022$)}          & \multicolumn{1}{c|}{0.008 ($\pm 0.003$)}          & \multicolumn{1}{c|}{0.818 ($\pm 0.019$)}          & 0.710 ($\pm 0.026$)          & \multicolumn{1}{c|}{0.124 ($\pm 0.021$)}          & \multicolumn{1}{c|}{0.014 ($\pm 0.003$)}          & \multicolumn{1}{c|}{0.232 ($\pm 0.035$)}          & 0.153 ($\pm 0.027$)          \\
\multicolumn{1}{c|}{DSC++} & \multicolumn{1}{c|}{0.017 ($\pm 0.007$)}          & \multicolumn{1}{c|}{\textbf{0.007 ($\pm 0.003$)}} & \multicolumn{1}{c|}{\textbf{0.911 ($\pm 0.008$)}} & \textbf{0.841 ($\pm 0.013$)} & \multicolumn{1}{c|}{0.045 ($\pm 0.006$)}          & \multicolumn{1}{c|}{0.012 ($\pm 0.002$)}          & \multicolumn{1}{c|}{\textbf{0.429 ($\pm 0.041$)}} & \textbf{0.311 ($\pm 0.036$)} \\ \hline
\end{tabular}}
\end{table*}

The DSC++ achieved significantly better calibration scores compared to the DSC loss across both classes (Kidney: $p=\num{3e-8}$, Kidney tumour: $p=\num{2e-8}$). In contrast, there was no significant difference in calibration scores between the DSC++ loss and CE-based losses. In terms of segmentation quality, the DSC++ achieved the best performance with a DSC score of 0.911 and 0.429 for the kidney and kidney tumour segmentation respectively. The DSC score on the kidney tumour class using the DSC++ loss significantly outperformed the other loss functions (DSC: $p=\num{2e-6}$, CE: $p=\num{4e-7}$, Focal: $p=\num{0.0002}$).

Example segmentations using each loss function on the KiTS19 dataset is shown in Figure 4. The DSC++ loss produces accurate and well calibrated segmentations, for both kidney and kidney tumour class. The DSC loss produces false positive predictions with high confidence, most noticable with the kidney tumour class. The CE-based losses produce poor quality kidney tumour segmentation, with associated over-segmentation of the kidney.

\begin{figure*}[ht!]
    \centering
    \includegraphics[scale=0.27]{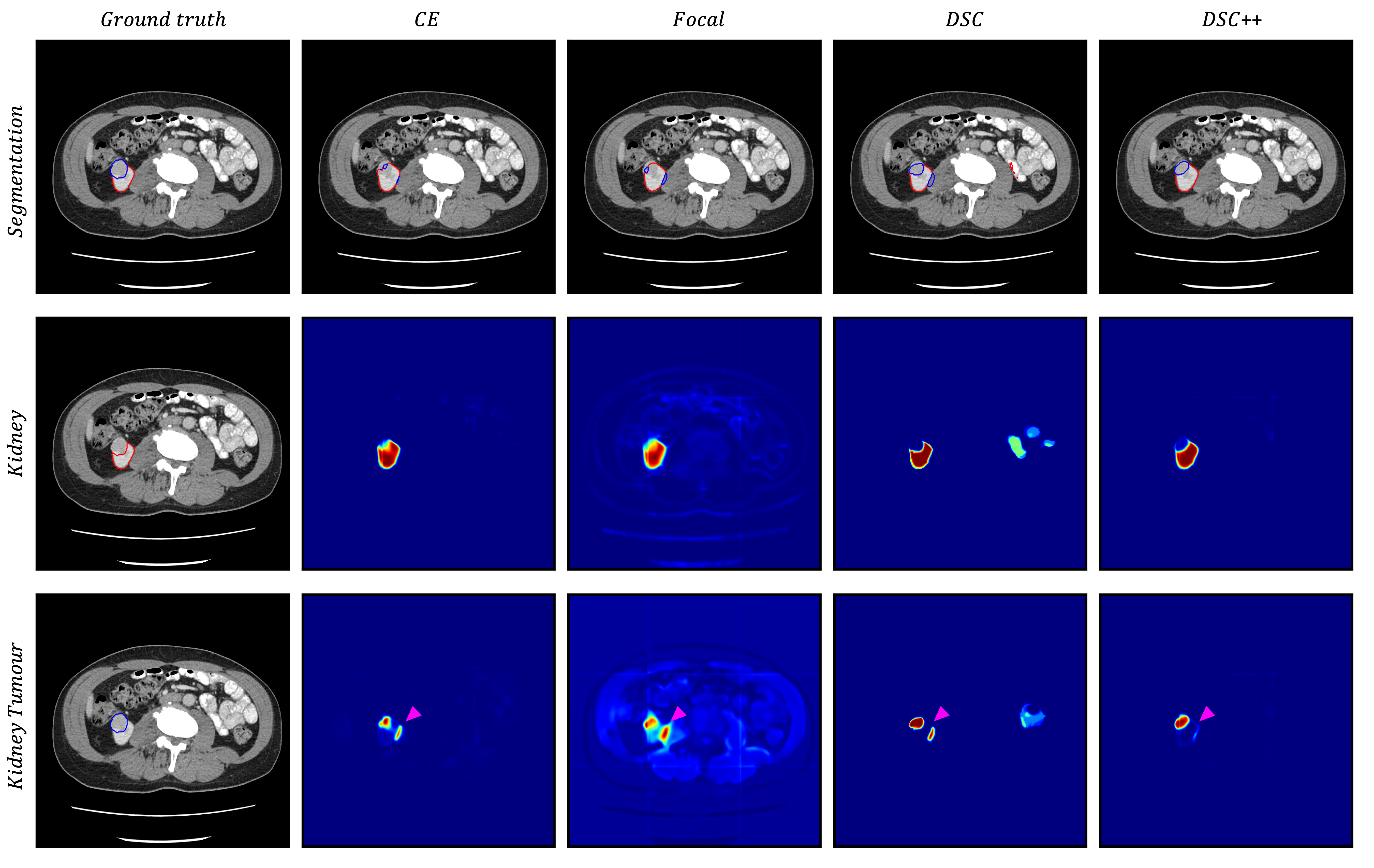}
    \caption{Example segmentation, with softmax predictions visualised as a heatmap, for each loss function. The image and ground truth are provided for reference. The pink arrows highlight example areas where segmentation quality differs.}
    \label{fig:figure_4}
\end{figure*}

\subsection{Incorporating the DSC++ loss into other Dice-based loss functions}

The DSC loss forms the basis for several other region-based loss functions, and therefore we investigate the effect of integrating the DSC++ loss modification into these loss functions. The results are shown in Table 6. 

\begin{table*}[ht!]
\centering
\caption{Calibration and performance of the DSC-based loss functions, using either the original loss functions (Tversky, Focal Tversky, Combo and Unified Focal) or substituting the DSC component of the loss for the DSC++ loss (Tversky++, Focal Tversky++, Combo++ and Unified Focal++). $\gamma$ is set to $2$ for the DSC++ variants. The standard errors are shown in brackets. The best scores are denoted in bold.}
\scalebox{0.75}{
\begin{tabular}{l|cc|cc}
\hline
                          & \multicolumn{2}{c|}{Calibration}                                     & \multicolumn{2}{c}{Performance}                                      \\ \hline
\multicolumn{1}{c|}{Loss} & \multicolumn{1}{c|}{NLL ($\downarrow$)}                    & Brier ($\downarrow$)                 & \multicolumn{1}{c|}{Dice ($\uparrow$)}                   & Jaccard ($\uparrow$)               \\ \hline
Tversky                   & \multicolumn{1}{c|}{0.144 ($\pm 0.011$)}          & 0.034 ($\pm 0.001$)          & \multicolumn{1}{c|}{0.807 ($\pm 0.003$)}          & 0.676 ($\pm 0.004$)          \\
Tversky++                 & \multicolumn{1}{c|}{\textbf{0.033 ($\pm 0.003)$}} & \textbf{0.025 ($\pm 0.001$)} & \multicolumn{1}{c|}{\textbf{0.810 ($\pm 0.003$)}} & \textbf{0.681 ($\pm 0.004$)} \\ \hline
Focal Tversky             & \multicolumn{1}{c|}{0.142 ($\pm 0.011$)}          & 0.033 ($\pm 0.001$)          & \multicolumn{1}{c|}{0.807 ($\pm 0.003$)}          & 0.677 ($\pm 0.004$)          \\
Focal Tversky++           & \multicolumn{1}{c|}{\textbf{0.036 ($\pm 0.003$)}} & \textbf{0.024 ($\pm 0.001$)} & \multicolumn{1}{c|}{\textbf{0.810 ($\pm 0.003$)}} & \textbf{0.680 ($\pm 0.004$)} \\ \hline
Combo                     & \multicolumn{1}{c|}{0.063 ($\pm 0.004$)}          & 0.025 ($\pm 0.001$)          & \multicolumn{1}{c|}{\textbf{0.802 ($\pm 0.004$)}} & 0.669 ($\pm 0.005$)          \\
Combo++                   & \multicolumn{1}{c|}{\textbf{0.050 ($\pm 0.003$)}} & \textbf{0.024 ($\pm 0.001$)} & \multicolumn{1}{c|}{\textbf{0.802 ($\pm 0.003$)}} & \textbf{0.670 ($\pm 0.005$)} \\ \hline
Unified Focal             & \multicolumn{1}{c|}{0.056 ($\pm 0.004$)}          & 0.026 ($\pm 0.001$)          & \multicolumn{1}{c|}{\textbf{0.810 ($\pm 0.003$)}} & 0.680 ($\pm 0.004$)          \\
Unified Focal++           & \multicolumn{1}{c|}{\textbf{0.039 ($\pm 0.003$)}} & \textbf{0.024 ($\pm 0.001$)} & \multicolumn{1}{c|}{\textbf{0.810 ($\pm 0.003$)}} & \textbf{0.681 ($\pm 0.004$)} \\ \hline
\end{tabular}}
\end{table*}

The DSC-based variants appear to all inherit the poorly calibrated nature of the DSC loss, except for the two compound loss functions, the Combo loss and the Unified Focal loss, which also incorporate the CE-based variants. Using the DSC++ loss led to significant improvements in calibration for all loss functions compared, for both the NLL and Brier scores. Similarly, the highest performance, measured using the DSC and Jaccard scores, were obtained using the DSC++ variants.

\subsection{Softmax thresholding}

The effect of softmax thresholding on the performance of the DSC and DSC++ loss for the DRIVE dataset are shown in Figure 5. The DSC loss predictions display almost no variation across the entire range of softmax thresholds. In contrast, there are significant variations in recall and precision scores using the DSC++ loss. Importantly, considerable increases in recall or precision between $\mathcal{T} = 0.3$ and $\mathcal{T} = 0.7$ did not affect the DSC score. The DSC++ loss enables models to be tailored to provide either very high recall or precision values, with little effect on the DSC score. For example, the model achieved a precision of $0.923$ and DSC of $0.748$ at $\mathcal{T}=0.8$, and recall of $0.923$ and DSC of $0.761$ at $\mathcal{T} = 0.2$.

\begin{figure*}[ht!]
    \centering
    \includegraphics[scale=0.25]{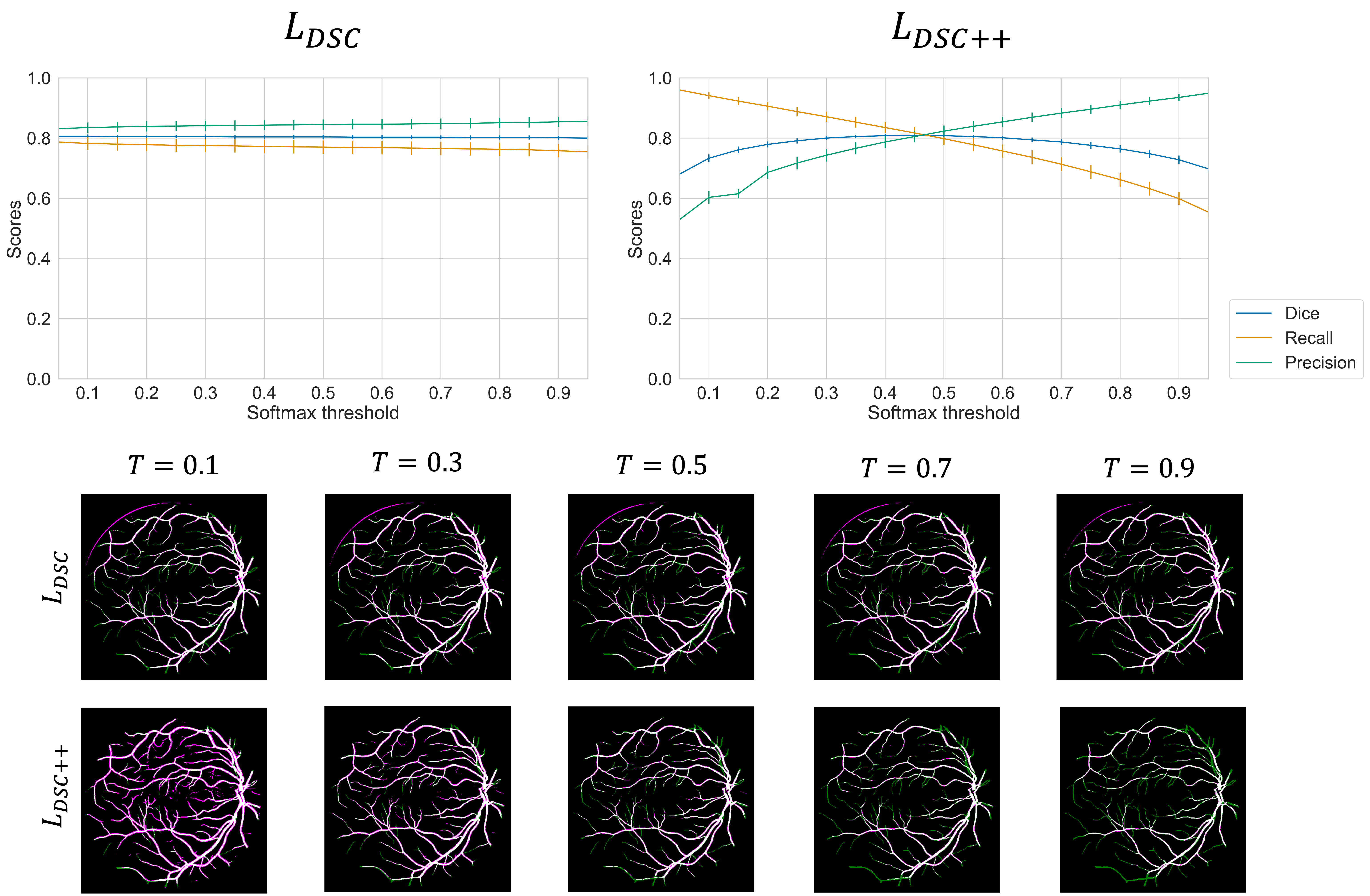}
    \caption{The effect of softmax thresholding on the recall and precision using models trained with the DSC and DSC++ loss on the DRIVE dataset. Top: Recall, precision and DSC scores at different softmax thresholds for the DSC and DSC++ loss. The vertical bars represent the $95\%$ confidence intervals. Bottom: Example segmentation output at different softmax thresholds. The false positive are highlighted in magenta, and the false negatives are highlighted in green.}
    \label{fig:figure_5}
\end{figure*}

\section{Discussion}

In this work, we identified the weights associated with the FP and FN predictions as the reason for the poor calibration associated with the DSC loss, and used this to provide a reformulation, known as the DSC++ loss, which uses a $\gamma$ parameter to more heavily penalise overconfident predictions. We observed significantly improved calibration using the DSC++ loss over the DSC loss, measured using the NLL and Brier scores, across six well-validated open-source datasets, including both 2D binary and 3D multi-class segmentation tasks. Furthermore, we demonstrated that the variants of the DSC loss inherit poor calibration, while those using DSC++ variants led to significant improvements in calibration. Finally, we evaluated the effect of softmax thresholding on the DSC loss and DSC++ loss, where little variation in recall or precision was observed with the DSC loss, in comparison to the significant variation achievable using the DSC++ loss.

Modifying the loss function, rather than the network or training setup, is the most intuitive solution to improve calibration. This is because with optimal training, it is the loss function that primarily determines the calibration quality of the resulting segmentation outputs. Optimisation with the DSC loss will encourage overconfident predictions (Figure 1), and therefore methods---such as MC dropout or deep ensembling---may improve calibration, but do not address the direct cause of the issue. Importantly, both MC dropout and deep ensembling significant increase inference time, with the latter requiring additional computational resources to handle predictions from multiple networks. Furthermore, MC dropout requires modifying networks to include dropout layers, and this may not be compatible with certain architectures.

We also explored the synergistic effect of softmax thresholding, together with well calibrated outputs, to enable tailoring towards high recall or high precision output states (Figure 5). For biomedical or clinical use, generally high recall is favoured, especially when the role of automatic segmentation systems is to support human operators in reducing false negative predictions, for example with polyp identification during colonoscopy \cite{nogueira2021deep}. As shown in Figure 5, it is possible to identify even the small diameter retinal vessels when recall is prioritised. It is possible to optimise models to produce high recall or precision outputs, such as the Tversky loss modification of the DSC loss \cite{salehi2017tversky}. However, after model training, it is not possible to further modify the recall-precision bias, which would instead require the training of a new model. Softmax thresholding is used during post-processing and is therefore independent of the model, enabling flexible and reversible control over the recall-precision bias. Even without softmax thresholding, the uncertainty associated with well calibrated predictions can highlight regions of interest which may be missed when interpreting poorly calibrated predictions (Figure 3 and 4).

Given the widespread use of these functions, it is important to consider whether there are any reasons to not replace them with these alternatives. The one apparent limitation of using the DSC++ loss over the DSC loss is additional hyperparameter tuning required. However, we investigated a large range of $\gamma$ values (Table III and Figure 2), and observed that performance was not significantly affected, while the calibration improves significantly, even with small values of $\gamma$. Moreover, we selected a $\gamma$ value of $2$ based on results from the DRIVE dataset, and this appeared to generalise well across the other five datasets, with consistently significant improvements to calibration (Table IV and V). Therefore, even small $\gamma$ parameter values appear to be effective, and optimal choices for $\gamma$ generalise well across datasets, suggesting that the $\gamma$ parameter is relatively easy to optimise.

It is less clear whether the DSC++ loss should be favoured above other loss functions. Besides calibration, the DSC++ loss suffers from the same limitations as the DSC loss, namely the unstable gradient, resulting from gradient calculations involving small denominators \cite{wong20183d, bertels2019optimization}. While there is currently little empirical evidence relating the unstable gradient to suboptimal performance, it has been suggested that incorporating the CE loss helps to mitigate the unstable gradients generated by the DSC loss \cite{isensee2021nnu}. Our experiments confirm previous results that compound loss functions generally perform better \cite{yeung2021unified,ma2021loss}. However, even if the DSC++ cannot replace these loss functions, we have shown that replacing the DSC component of loss functions with the DSC++ loss leads to significant improvements in calibration, as well as evidence of better performance (Table VI). 

In future work, we will investigate the effect of gradient instability on the performance of the DSC++ loss. It would be important to evaluate the performance on highly class imbalanced datasets, where gradient stabilisation may be expected to be more important. Furthermore, it would be useful to evaluate networks trained using the DSC++ loss on out-of-distribution data, to test whether the model predictions remain well calibrated.

\section{Conclusion}

In this study, we identified the main reason behind neural network overconfidence when training deep learning-based image segmentation models using the DSC loss, and provided a simple yet effective modification, named the DSC++ loss, that directly addresses the issue. After evaluating the performance and calibration of both the DSC loss and DSC++ loss across six well-validated biomedical imaging datasets, as well as systematically analysing the softmax predictions, it is clear that the DSC loss is not suitable for training neural networks for use in biomedical or clinical practice. In contrast, the DSC++ loss, together with its synergistic effect using softmax thresholding, produce model outputs that are useful to interpret, and readily adjustable to provide high recall or precision outputs. Compared with previous methods used to improve the calibration of networks trained using the DSC loss, the DSC++ loss provides the most intuitive, readily accessible solution that is an important contribution towards the goal of deploying deep learning image segmentation systems into biomedical or clinical practice.

\pagebreak

\section*{Statements and Declarations}

\subsection{Funding}

This work was partially supported by The Mark Foundation for Cancer Research and Cancer Research UK Cambridge Centre [C9685/A25177], the CRUK National Cancer Imaging Translational Accelerator (NCITA) [C42780/A27066] and the Wellcome Trust Innovator Award, UK [215733/Z/19/Z].
Additional support was also provided by the National Institute of Health Research (NIHR) Cambridge Biomedical Research Centre [BRC-1215-20014] and the Cambridge Mathematics of Information in Healthcare (CMIH) [funded by the EPSRC grant EP/T017961/1].
The views expressed are those of the authors and not necessarily those of the NHS, the NIHR, or the Department of Health and Social Care.

This work was supported in part by the UK Research and Innovation Future Leaders Fellowship [MR/V023799/1], in part by the Medical Research Council [MC/PC/21013], in part by the European Research Council Innovative Medicines Initiative [DRAGON, H2020-JTI-IMI2 101005122], and in part by the AI for Health Imaging Award [CHAIMELEON, H2020-SC1-FA-DTS2019-1 952172]. 

CBS in addition acknowledges support from the Leverhulme Trust project on `Breaking the non-convexity barrier', the Philip Leverhulme Prize, the Royal Society Wolfson Fellowship, the EPSRC grants EP/S026045/1, EP/N014588/1, European Union Horizon 2020 research and innovation programmes under the Marie Skodowska-Curie grant agreement No. 777826 NoMADS and No. 691070 CHiPS, the Cantab Capital Institute for the Mathematics of Information and the Alan Turing Institute.

This work was performed using resources provided by the Cambridge Service for Data Driven Discovery (CSD3) operated by the University of Cambridge Research Computing Service (\url{www.csd3.cam.ac.uk}), provided by Dell EMC and Intel using Tier-2 funding from the Engineering and Physical Sciences Research Council (capital grant EP/P020259/1), and DiRAC funding from the Science and Technology Facilities Council (\url{www.dirac.ac.uk}).

\subsection{Competing Interests}

The authors have no relevant financial or non-financial interests to disclose.

\bigskip
\begin{flushleft}%
Editorial Policies for:

\bigskip\noindent
Springer journals and proceedings: \url{https://www.springer.com/gp/editorial-policies}

\bigskip\noindent
Nature Portfolio journals: \url{https://www.nature.com/nature-research/editorial-policies}

\bigskip\noindent
\textit{Scientific Reports}: \url{https://www.nature.com/srep/journal-policies/editorial-policies}

\bigskip\noindent
BMC journals: \url{https://www.biomedcentral.com/getpublished/editorial-policies}
\end{flushleft}

\bibliography{sn-bibliography}


\end{document}